\documentclass[prl,aps,twocolumn,floatfix]{revtex4}

\usepackage{graphicx,epsf,subfigure,amsbsy,amsmath,float}
\usepackage[latin1]{inputenc}

\newcommand{\ket}[1]{| #1 \rangle}

\newcommand{\bracket}[3]{\langle #1 | #2 | #3 \rangle}

\begin{document}

\bibliographystyle{apsrev}

\title{Identification of the dominant precession damping mechanism in Fe, Co, and Ni by 
first-principles calculations}

\author{K. Gilmore$^{1,2}$, Y.U. Idzerda$^2$, and M.D. Stiles$^1$} 

\affiliation{$^1$ National Institute of Standards and Technology, Gaithersburg, 
MD 20899-8412 \\
$^2$ Physics Department, Montana State University, Bozeman, MT 59717}

\date{\today}

\begin{abstract}

The Landau-Lifshitz equation reliably describes magnetization dynamics using a 
phenomenological treatment of damping.  This paper presents first-principles 
calculations of the damping parameters for Fe, Co, and Ni that quantitatively 
agree with existing ferromagnetic resonance measurements.  This agreement establishes the dominant 
damping mechanism for these systems and takes a significant step toward predicting and tailoring the 
damping constants of new materials.
\end{abstract}

\pacs{PACS numbers: }

\maketitle



Magnetic damping determines the performance of magnetic devices including hard drives, 
magnetic random access memories, magnetic logic devices, and magnetic field sensors.  
The behavior of these devices can be modeled using the Landau-Lifshitz (LL) equation 
\cite{Landau.Lifshitz:1935}
~
\begin{equation}
\dot{{\bf m}} = -|\gamma| {\bf m} \times {\bf H}_{\rm eff} - \frac{\lambda}{m^2} 
{\bf m} \times \left ( {\bf m} \times {\bf H}_{\rm eff} \right ),
\label{ll_equation}
\end{equation}

\noindent or the essentially equivalent Gilbert (LLG) form \cite{Gilbert:1956, 
Gilbert:2004}.  The first term describes precession of the magnetization ${\bf m}$ about the 
effective field ${\bf H}_{\rm 
eff}$ where $\gamma = g\mu_0\mu_B/\hbar$ is the gyromagnetic ratio.  The second term is a 
phenomenological 
treatment of damping with the adjustable rate $\lambda$.  The LL(G) equation adequately 
describes dynamics measured by techniques as varied as ferromagnetic resonance (FMR) 
\cite{McMichael:2003b}, magneto-optical Kerr effect \cite{Rasing:2001}, x-ray absorption 
spectroscopy \cite{Bailey:2004}, and spin-current driven rotation with the addition of a 
spin-torque term \cite{Krivorotov:2007, Stiles:2006}.

Access to a range of damping rates in metallic materials is desirable when constructing 
devices for different applications.  Ideally, one would like the ability to design 
materials with any desired damping rate.  Empirically, doping NiFe alloys with 
transition metals \cite{McMichael:2007} or rare earths \cite{Bailey:2001} has produced 
compounds with damping rates in the range of $\alpha = 0.01$ to 0.8.  A recent 
investigation of adding vanadium to iron resulted in an alloy with a damping rate slightly 
lower than that for pure iron \cite{Bailey:2007}, the system with the lowest previously 
known value.  However, the damping rate of a new material cannot be predicted because there has not yet 
been a first-principles calculation of 
damping that quantitatively agrees with experiment.  The challenging pursuit of new 
materials with specific or lowered damping rates is further complicated by the 
expectation that, as device size continues to be scaled down, material parameters, 
such as $\lambda$, should change \cite{Fahnle:2005}.  A detailed understanding of the 
important damping mechanisms in metallic ferromagnets and the ability to predictively 
calculate damping rates would greatly facilitate the design of new materials appropriate for a 
variety of applications.

The temperature dependence of damping in the transition metals has been carefully characterized 
through measurement of small angle dynamics by FMR \cite{Bhagat.Lubitz:1974}.  While one might 
na\"{i}vely expect damping to increase monotonically with temperature, as it does for Fe, both Co 
and Ni also exhibit a dramatic rise in damping at low temperature as the temperature decreases.  
These observations indicate that two primary mechanisms are involved.  Subsequent experiments 
\cite{Heinrich:1979, Heinrich:1980} partition these non-monotonic damping curves into a {\em 
conductivity-like} term that decreases with temperature and a {\em resistivity-like} term that 
increases with temperature.  The two terms were found to give nearly equal weight to the
damping curve of Ni and have temperature dependencies similar to those of the conductivity and 
resistivity, suggesting two distinct roles for electron-lattice scattering.

The torque-correlation model of Kambersky \cite{Kambersky:1976} appears to qualitatively match the 
data.  However, like most of the various models presented by Kambersky \cite{Kambersky:1967, 
Kambersky:1970, Kambersky:1976, Kambersky:1984} and others \cite{Korenman.Prange:1972}, it has not 
been quantitatively evaluated in a rigorous fashion.  This 
has left the community to speculate, based on rough estimates or less, as to which damping 
mechanisms are important.  We resolve this matter in the present work by reporting first-principles 
calculations of the Landau-Lifshitz damping constant according to Kambersky's torque-correlation 
expression.  Quantitative comparison of the present calculations to the measured FMR values
\cite{Bhagat.Lubitz:1974} positively identifies this damping pathway as the dominant effect in the 
transition metal systems.  In addition to presenting these primary conclusions, we also
describe the relationship between the torque-correlation model and the more widely understood 
breathing Fermi surface model \cite{Kambersky:1970, Kambersky:2002}, showing that the 
results of both models agree quantitatively in the low scattering rate limit.

The breathing Fermi surface model of Kambersky predicts
~
\begin{equation}
\lambda = \frac{g^2\mu_{\rm B}^2}{\hbar} \sum_{n} \int \frac{dk^3}{(2\pi)^3} \eta(\epsilon_{n,k}) \left ( 
\frac{\partial \epsilon_{n,k}}{\partial \theta} \right )^2  \frac{\tau}{\hbar} \, .
\label{BFS}
\end{equation}

\noindent This model offers a qualitative explanation for the low 
temperature {\em conductivity-like} contribution to the measured damping.  The model 
describes damping of uniform precession as due to variations $\partial \epsilon_{n,k}/\partial 
\theta$ in the energies $\epsilon_{n,k}$ of the single-particle states with respect to the spin 
direction $\theta$.  The states are labeled with a wavevector $k$ and band index $n$.  As 
the magnetization precesses, the spin-orbit interaction changes the energy of 
electronic states pushing some occupied states above the Fermi level and some unoccupied states 
below the Fermi level.  Thus, electron-hole pairs are generated near the Fermi level even in the 
absence of changes in the electronic populations.  The $\eta$ 
function in Eq.~(\ref{BFS}) is the negative derivative of the Fermi function and picks out 
only states near the Fermi level to contribute to the damping.  $g$ is the Land\'{e} g-factor and 
$\mu_{\rm B}$ is the Bohr magneton.  The electron-hole 
pairs created by the precession exist for some lifetime $\tau$ before relaxing through lattice 
scattering.  The amount of energy and angular momentum dissipated to the lattice depends on how 
far from equilibrium the system gets, thus damping by this mechanism increases linearly with 
the electron lifetime as seen in Eq.\ref{BFS}.  Since the electron lifetime is expected to 
decrease as the temperature increases, this model predicts that damping diminishes as the 
temperature is raised.

Because the predicted damping rate is linear in the scattering time the damping rate cannot be  
calculated more accurately than the scattering time is known.  For this reason it is not possible to 
make quantitative comparisons between calculations of the breathing Fermi surface and measurements.
Further, while the breathing Fermi surface model can explain the dramatic temperature dependence observed 
in the {\em conductivity-like} portion of the data it fails to capture the 
physics driving the {\em resistivity-like} term.  This is a significant limitation from a practical 
perspective because the {\em resistivity-like} term dominates damping at room temperature and 
above and is the only contribution observed in iron \cite{Bhagat.Lubitz:1974} and NiFe alloys 
\cite{Ingvarsson:2002}.  For these reasons it is necessary to turn to more complete models of damping. 

Kambersky's torque-correlation model predicts 
~
\begin{equation}
\lambda = \frac{g^2\mu_B^2}{\hbar} \sum_{n,m} \int \frac{dk^3}{(2\pi)^3} \, \left |
\Gamma^-_{nm}(k) \right |^2 W_{nm}(k) 
\label{lambda}
\end{equation}

\noindent and we will show that it both incorporates the physics of the breathing Fermi surface 
model and also accounts for the {\em resistivity-like} terms.  The matrix elements $\Gamma^-_{nm}(k) = 
\bracket{n,k}{[\sigma^- \, , \, H_{\rm so}]}{m,k}$ measure transitions between states in bands 
$n$ and $m$ induced by the spin-orbit torque.  These transitions conserve wavevector $k$ 
because they describe the annihilation of a uniform precession magnon, which carries no linear 
momentum.  The nature of these scattering events, which are weighted by the spectral overlap $W_{nm}(k) = 
(1/\pi)\int d\omega_1 \, \eta(\omega_1) A_{nk}(\omega_1) A_{mk}(\omega_1)$, will be discussed in more 
detail below.  The electron spectral functions $A_{nk}$ are Lorentzians centered around the 
band energies $\epsilon_{nk}$ and broadened by interactions with the lattice.  The width of the 
spectral function $\hbar/\tau$ provides a phenomenological account for the role of 
electron-lattice scattering in the damping process.  The $\eta$ function is the same as in 
Eq.~(\ref{BFS}) and enforces the requirement of spectral overlap at the Fermi level.

\begin{figure}
\begin{center}
\includegraphics[angle=0,width=8.0cm]{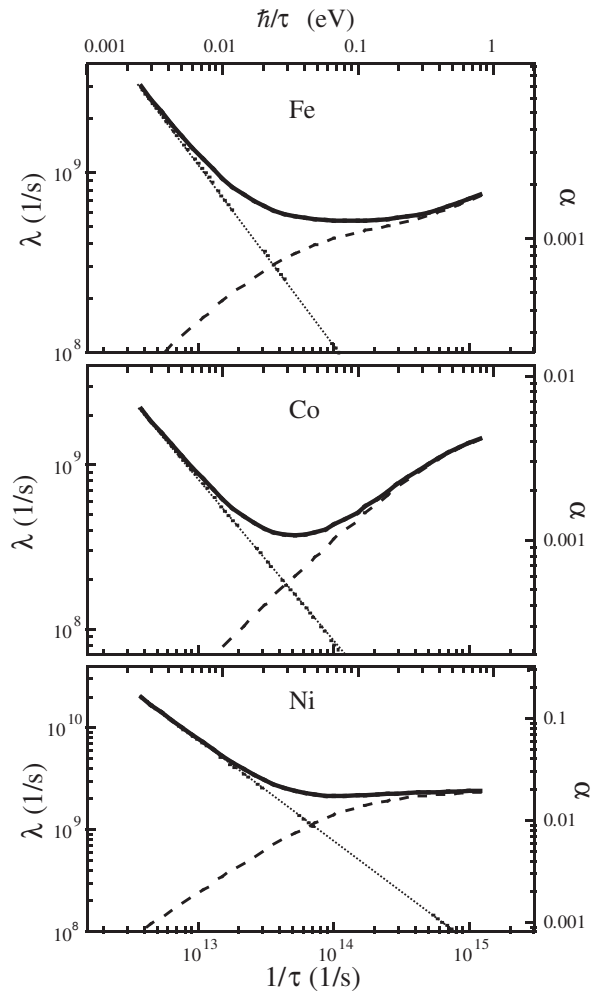}
\end{center}
\caption{Calculated Landau-Lifshitz damping constant for Fe, Co, and Ni.  Thick solid
curves give the total damping parameter while dotted curves give the intraband and dashed
lines the interband contributions.  The top axis is the full-width-half-maximum of the electron spectral
functions. }
\label{calculations}
\end{figure}

\begin{table*}[tbh]
  \caption{Calculated and measured \cite{Bhagat.Lubitz:1974} damping parameters.  Values
for $\lambda$, the Landau-Lifshitz form, are reported in $10^{9} \, s^{-1}$, values of
$\alpha$, the Gilbert form, are dimensionless.  The last two columns list calculated
damping due to the intraband contribution from Eq.~(\ref{lambda}) and from the breathing
Fermi surface model \cite{Fahnle:2005}, respectively.  Values for $\lambda/\tau$ are given
in $10^{22} \, s^{-2}$.  Published numbers from \cite{Bhagat.Lubitz:1974} and
\cite{Fahnle:2005} have been multiplied by $4\pi$ to convert from the cgs unit system to
SI.}
\medskip
\begin{center}
\begin{tabular}{lcccccc}  \hline \hline
  & \quad \quad $\alpha_{\rm calc}$ \quad & \quad
$\lambda_{\rm calc}$ \quad & \quad $\lambda_{\rm meas}$ \quad &
\quad $\lambda_{\rm calc}/\lambda_{\rm meas}$ \quad & \quad $(\lambda/\tau)_{\rm intra}$
\quad & \quad $(\lambda/\tau)_{\rm BFS}$ \quad \\  \hline
bcc Fe $\langle 001 \rangle$   & 0.0013 & 0.54  & 0.88 & 0.61 & 1.01 & 0.968 \\
bcc Fe $\langle 111 \rangle$   & 0.0013 & 0.54  & --   & --   & 1.35 & 1.29  \\
hcp Co $\langle 0001 \rangle$  & 0.0011 & 0.37  & 0.9  & 0.41 & 0.786& 0.704 \\
fcc Ni $\langle 111 \rangle$   & 0.017  & 2.1   & 2.9  & 0.72 & 6.67 & 6.66  \\
fcc Ni $\langle 001 \rangle$   & 0.018  & 2.2   & --   & --   & 8.61 & 8.42  \\
\hline  \hline
\end{tabular}
\end{center}
\label{results}
\end{table*}

Equation (\ref{lambda}) captures two different types of scattering
events: scattering within a single band, $m$$\,=\,$$n$, for which the
initial and final states are the same, and scattering between two
different bands, $m$$\,\neq\,$$n$.  As explained in
\cite{Kambersky:1976} the overlap of the spectral functions is
proportional (inverse) to the electron scattering time for intraband
(interband) scattering.  From this observation the qualitative
conclusion is made that the intraband contributions match the {\em
conductivity-like} terms while the interband contributions give the
{\em resistivity-like} terms.  While this seems promising, evaluation
of Eq.~(\ref{lambda}) is more computationally intensive than that of
the breathing Fermi surface model and until now only a few estimates
for Ni and Fe have been made \cite{Kambersky:1984}.

We have performed first-principles calculations of the
torque-correlation model Eq.~(\ref{lambda}) with realistic band
structures for Fe, Co, and Ni.  Prior to evaluating Eq.~(\ref{lambda})
the eigenstates and energies of each metal were found using the linear
augmented plane wave method \cite{Mattheiss.Hamann:1986} in the local spin density approximation (LSDA)
\cite{Hohenberg.Kohn:1964, Kohn.Sham:1965, Barth.Hedin:1972}.  Details
of the calculations for these materials are described in
\cite{Stiles:2001}.  The exchange field was fixed in the chosen
equilibrium magnetization direction.  Calculations of
Eq.~(\ref{lambda}) presented in this paper are converged to within a
standard deviation of 3 $\%$, which required sampling $(160)^3$
$k$-points for Fe, $(120)^3$ for Ni, and $(100)^2$ $k$-points in
the basal plane by 57 along the $c$-axis for Co.  Electron-lattice
interactions were treated phenomenologically as a broadening of the
spectral functions.  The Fermi distribution was smeared with an
artificial temperature.  Results did not vary significantly with
reasonable choices of this temperature since the broadening of the
Fermi distribution was considerably less than that of the bands.  The
damping rate was calculated for a range of scattering rates (spectral
widths) just as damping has been measured over a range of
temperatures.

The results of these calculations are presented in
Fig.~\ref{calculations} and are decomposed into the intraband and
interband terms.  
The downward sloping line in Fig.~\ref{calculations} represents the
intraband contribution to damping.  Damping constants were recently
calculated using the breathing Fermi surface model \cite{Fahnle:2005,
Kambersky:2002} by evaluating the derivative of the electronic energy
with respect to the spin direction according to Eq.~(\ref{BFS}).  
The results of the breathing Fermi surface
prediction are indistinguishable from the intraband terms of the
present calculation even though the computational approaches differed
significantly; the agreement is quantified in Table \ref{results}.

The breathing Fermi surface model could not be quantitatively compared to the experimental
results because the temperature dependence of the scattering rate has not been determined
sufficiently accurately.  While the
present calculations also require knowledge of the scattering rate to determine the damping
rate the non-monotonic dependence of damping on the scattering rate produces a unique minimum
damping rate.  In the same manner that the calculated curves of Fig.~\ref{calculations} have a
minimum with respect to scattering rate, the measured damping curves exhibit minima with respect to
temperature.  Whatever the relation between temperature and scattering rate, the calculated minima
may be compared directly and quantitatively to the measured minima.  Table \ref{results} makes this
comparison.  The agreement between measured and calculated values shows that the torque-correlation model 
accounts for the dominant contribution to damping in these systems.

Our calculated values are smaller than the measured values.  Using
measured $g$ values instead of setting $g$ = $2$ would increase our results by a
factor of $(g/2)^2$, or about 10 $\%$ for Fe and 20 $\%$ for Co and Ni.
Other possible reasons for the difference include a simplified treatment of
electron-lattice scattering in which the scattering rates for all states were assumed
equal, the mean-field approximation for the exchange interaction, errors associated with the
local spin density approximation (LSDA), and numerical convergence (discussed below).  Other
damping mechanisms may also make small contributions \cite{McMichael:2002, Heinonen:2005,
Tserkovnyak:2004}.

Since the manipulations involved with the equation
of motion techniques employed in deriving Eq.~(\ref{lambda}) obscure
the underlying physics we now discuss the two scattering processes and
connect the intraband terms to the breathing Fermi surface model.
The intraband terms in Eq.~(\ref{lambda}) describe scattering from one
state to itself by the torque operator, which is similar to a
spin-flip operator.  A spin-flip operation between some state and
itself is only non-zero because the spin-orbit interaction mixes small
amounts of the opposite spin direction into each state.  Since the
initial and final states are the same, the operation is naturally spin
conserving.  The matrix elements do not describe a real transition,
but rather provide a measure of the energy of the electron-hole pairs
that are generated as the spin direction changes.  The electron-hole
pairs are subsequently annihilated by a real electron-lattice
scattering event.

To connect the derivatives $\partial\epsilon/\partial\theta$ in Eq.~(\ref{BFS}) and the torque matrix 
elements in Eq.~(\ref{lambda}) we imagine first pointing the magnetization in some direction $\hat{z}$.  
The only energy that changes with the magnetization direction is the spin-orbit energy $H_{\rm so}$.  As 
the spin of a single particle state $\ket{}$ rotates along $\hat{\theta}$ about $\hat{x}$ 
its spin-orbit energy is given by $\epsilon(\theta) = \bracket{}{e^{i\sigma_x \theta} \, H_{\rm so} \, 
e^{-i\sigma_x \theta}}{}$.  The derivative with respect to $\theta$ is 
$\partial\epsilon(\theta)/\partial\theta = i\bracket{}{e^{i\sigma_x \theta} [\sigma_x \, , \, H_{\rm so}] 
e^{-i\sigma_x \theta}}{}$.  Evaluating this derivative at the pole ($\theta$ = 0) gives 
$\partial\epsilon/\partial\theta = i\bracket{}{[\sigma_x \, , \, H_{\rm so}]}{}$.  Similarly, rotating 
the spin along $\hat{\theta}$ about $\hat{y}$ leads to $\partial\epsilon/\partial\theta = 
i\bracket{}{[\sigma_y \, , \, H_{\rm so}]}{}$.  The torque matrix elements in Eq.~(\ref{lambda}) are 
$\Gamma^- = \bracket{}{[\sigma^- \, , \, H_{\rm so}]}{} = \bracket{}{[\sigma_x \, , \, H_{\rm so}]}{} - 
i\bracket{}{[\sigma_y \, , \, H_{\rm so}]}{}$.  Using the relations between the commutators and 
derivatives just found the torque 
is $\Gamma^- = -i(\partial\epsilon/\partial\theta)_x - (\partial \epsilon/\partial \theta)_y$ where the 
subscripts indicate the rotation axis.  Squaring the torque matrix elements gives $|\Gamma^-|^2 = 
(\partial \epsilon/\partial \theta )_x^2 + (\partial \epsilon/ \partial \theta )_y^2$.  For high symmetry 
directions $(\partial \epsilon/\partial \theta)_x = (\partial \epsilon/ \partial \theta)_y$ and we deduce 
$|\Gamma^-|^2 = 2(\partial\epsilon/\partial \theta )^2$ demonstrating that the intraband terms 
of the torque-correlation model describe the same physics as the breathing Fermi surface.

The monotonically increasing curves in Fig.~\ref{calculations} indicate the interband 
contribution to damping.  Uniform mode magnons, which have negligible energy, may induce
quasi-elastic transitions between states with different energies.  This occurs when lattice
scattering broadens bands sufficiently so that they overlap at the Fermi level.  These
wavevector conserving transitions, which are driven by the precessing exchange field, occur primarily 
between states with significantly different spin character.  The process may
roughly be thought of as the decay of a uniform precession magnon into a single electron 
spin-flip excitation.  These events occur more frequently as the band overlaps increase.  For 
this reason the interband terms, which qualitatively match the {\em resistivity-like} 
contributions in the experimental data, dominate damping at room temperature and above.  

We have calculated the Landau-Lifshitz damping parameter for the itinerant ferromagnets Fe,
Co, and Ni as a function of the electron-lattice scattering rate.  The intraband and interband 
components match qualitatively to {\em conductivity-} and {\em resistivity-like} terms observed in 
FMR measurements.  A quantitative comparison 
was made between the minimal damping rates calculated as a function of scattering rate and measured 
with respect to temperature.  This comparison demonstrates that our calculations account for the 
dominant contribution to damping in these systems and identify the primary damping mechanism.  At 
room temperature and above damping occurs overwhelmingly through the interband transitions.  The 
contribution of these terms depends in part on the band gap spectrum around the Fermi level, which 
could be adjusted through doping.

K.G. and Y.U.I. acknowledge the support of the Office of Naval Research through grant 
N00014-03-1-0692 and through grant N00014-06-1-1016.  We would like 
to thank R.D. McMichael and T.J. Silva for valuable discussions.

\bibliography{all.refs}

\begin{thebibliography}{30}
\expandafter\ifx\csname natexlab\endcsname\relax\def\natexlab#1{#1}\fi
\expandafter\ifx\csname bibnamefont\endcsname\relax
  \def\bibnamefont#1{#1}\fi
\expandafter\ifx\csname bibfnamefont\endcsname\relax
  \def\bibfnamefont#1{#1}\fi
\expandafter\ifx\csname citenamefont\endcsname\relax
  \def\citenamefont#1{#1}\fi
\expandafter\ifx\csname url\endcsname\relax
  \def\url#1{\texttt{#1}}\fi
\expandafter\ifx\csname urlprefix\endcsname\relax\def\urlprefix{URL }\fi
\providecommand{\bibinfo}[2]{#2}
\providecommand{\eprint}[2][]{\url{#2}}

\bibitem[{\citenamefont{Landau and Lifshitz}(1935)}]{Landau.Lifshitz:1935}
\bibinfo{author}{\bibfnamefont{L.}~\bibnamefont{Landau}} \bibnamefont{and}
  \bibinfo{author}{\bibfnamefont{E.}~\bibnamefont{Lifshitz}},
  \bibinfo{journal}{Phys.~Z.~Sowjet.} \textbf{\bibinfo{volume}{8}},
  \bibinfo{pages}{153} (\bibinfo{year}{1935}).

\bibitem[{\citenamefont{Gilbert}(1956)}]{Gilbert:1956}
\bibinfo{author}{\bibfnamefont{T.~L.} \bibnamefont{Gilbert}},
  \bibinfo{journal}{Armour research foundation project No. A059, supplementary
  report, unpublished}  (\bibinfo{year}{1956}).

\bibitem[{\citenamefont{Gilbert}(2004)}]{Gilbert:2004}
\bibinfo{author}{\bibfnamefont{T.~L.} \bibnamefont{Gilbert}},
  \bibinfo{journal}{IEEE Trans.~Magn.} \textbf{\bibinfo{volume}{40}},
  \bibinfo{pages}{3443} (\bibinfo{year}{2004}).

\bibitem[{\citenamefont{Twisselmann and McMichael}(2003)}]{McMichael:2003b}
\bibinfo{author}{\bibfnamefont{D.}~\bibnamefont{Twisselmann}} \bibnamefont{and}
  \bibinfo{author}{\bibfnamefont{R.}~\bibnamefont{McMichael}},
  \bibinfo{journal}{J.~Appl.~Phys.} \textbf{\bibinfo{volume}{93}},
  \bibinfo{pages}{6903} (\bibinfo{year}{2003}).

\bibitem[{\citenamefont{Gerrits et~al.}(2001)\citenamefont{Gerrits, Hohlfeld,
  Gielkens, Veenstra, Bal, Rasing, and van~den Berg}}]{Rasing:2001}
\bibinfo{author}{\bibfnamefont{T.}~\bibnamefont{Gerrits}},
  \bibinfo{author}{\bibfnamefont{J.}~\bibnamefont{Hohlfeld}},
  \bibinfo{author}{\bibfnamefont{O.}~\bibnamefont{Gielkens}},
  \bibinfo{author}{\bibfnamefont{K.}~\bibnamefont{Veenstra}},
  \bibinfo{author}{\bibfnamefont{K.}~\bibnamefont{Bal}},
  \bibinfo{author}{\bibfnamefont{T.}~\bibnamefont{Rasing}}, \bibnamefont{and}
  \bibinfo{author}{\bibfnamefont{H.}~\bibnamefont{van~den Berg}},
  \bibinfo{journal}{J.~Appl.~Phys.} \textbf{\bibinfo{volume}{89}},
  \bibinfo{pages}{7648} (\bibinfo{year}{2001}).

\bibitem[{\citenamefont{Bailey et~al.}(2004)\citenamefont{Bailey, Cheng,
  Keavney, Kao, Vescovo, and Arena}}]{Bailey:2004}
\bibinfo{author}{\bibfnamefont{W.}~\bibnamefont{Bailey}},
  \bibinfo{author}{\bibfnamefont{L.}~\bibnamefont{Cheng}},
  \bibinfo{author}{\bibfnamefont{D.}~\bibnamefont{Keavney}},
  \bibinfo{author}{\bibfnamefont{C.}~\bibnamefont{Kao}},
  \bibinfo{author}{\bibfnamefont{E.}~\bibnamefont{Vescovo}}, \bibnamefont{and}
  \bibinfo{author}{\bibfnamefont{D.}~\bibnamefont{Arena}},
  \bibinfo{journal}{Phys.~Rev.~B} \textbf{\bibinfo{volume}{70}},
  \bibinfo{pages}{172403} (\bibinfo{year}{2004}).

\bibitem[{\citenamefont{Krivorotov et~al.}(2007)\citenamefont{Krivorotov,
  Berkov, Gorn, Emley, Sankey, Ralph, and Buhrman}}]{Krivorotov:2007}
\bibinfo{author}{\bibfnamefont{I.}~\bibnamefont{Krivorotov}},
  \bibinfo{author}{\bibfnamefont{D.}~\bibnamefont{Berkov}},
  \bibinfo{author}{\bibfnamefont{N.}~\bibnamefont{Gorn}},
  \bibinfo{author}{\bibfnamefont{N.}~\bibnamefont{Emley}},
  \bibinfo{author}{\bibfnamefont{J.}~\bibnamefont{Sankey}},
  \bibinfo{author}{\bibfnamefont{D.}~\bibnamefont{Ralph}}, \bibnamefont{and}
  \bibinfo{author}{\bibfnamefont{R.}~\bibnamefont{Buhrman}},
  \bibinfo{journal}{Phys.~Rev.~B}  (\bibinfo{year}{2007}).

\bibitem[{\citenamefont{Stiles and Miltat}(2006)}]{Stiles:2006}
\bibinfo{author}{\bibfnamefont{M.}~\bibnamefont{Stiles}} \bibnamefont{and}
  \bibinfo{author}{\bibfnamefont{J.}~\bibnamefont{Miltat}},
  \emph{\bibinfo{title}{Spin dynamics in confined magnetic structures III}}
  (\bibinfo{publisher}{Springer, Berlin}, \bibinfo{year}{2006}).

\bibitem[{\citenamefont{Rantschler et~al.}(2007)\citenamefont{Rantschler,
  McMichael, Castiello, Shapiro, W.F.~Egelhoff, Maranville, Pulugurtha, Chen,
  and Conners}}]{McMichael:2007}
\bibinfo{author}{\bibfnamefont{J.}~\bibnamefont{Rantschler}},
  \bibinfo{author}{\bibfnamefont{R.}~\bibnamefont{McMichael}},
  \bibinfo{author}{\bibfnamefont{A.}~\bibnamefont{Castiello}},
  \bibinfo{author}{\bibfnamefont{A.}~\bibnamefont{Shapiro}},
  \bibinfo{author}{\bibfnamefont{J.}~\bibnamefont{W.F.~Egelhoff}},
  \bibinfo{author}{\bibfnamefont{B.}~\bibnamefont{Maranville}},
  \bibinfo{author}{\bibfnamefont{D.}~\bibnamefont{Pulugurtha}},
  \bibinfo{author}{\bibfnamefont{A.}~\bibnamefont{Chen}}, \bibnamefont{and}
  \bibinfo{author}{\bibfnamefont{L.}~\bibnamefont{Conners}},
  \bibinfo{journal}{J.~Appl.~Phys.} \textbf{\bibinfo{volume}{101}},
  \bibinfo{pages}{033911} (\bibinfo{year}{2007}).

\bibitem[{\citenamefont{Bailey et~al.}(2001)\citenamefont{Bailey, Kabos,
  Mancoff, and Russek}}]{Bailey:2001}
\bibinfo{author}{\bibfnamefont{W.}~\bibnamefont{Bailey}},
  \bibinfo{author}{\bibfnamefont{P.}~\bibnamefont{Kabos}},
  \bibinfo{author}{\bibfnamefont{F.}~\bibnamefont{Mancoff}}, \bibnamefont{and}
  \bibinfo{author}{\bibfnamefont{S.}~\bibnamefont{Russek}},
  \bibinfo{journal}{IEEE Trans.~Mag.} \textbf{\bibinfo{volume}{37}},
  \bibinfo{pages}{1749} (\bibinfo{year}{2001}).

\bibitem[{\citenamefont{Scheck et~al.}(2007)\citenamefont{Scheck, Cheng,
  Barsukov, Frait, and Bailey}}]{Bailey:2007}
\bibinfo{author}{\bibfnamefont{C.}~\bibnamefont{Scheck}},
  \bibinfo{author}{\bibfnamefont{L.}~\bibnamefont{Cheng}},
  \bibinfo{author}{\bibfnamefont{I.}~\bibnamefont{Barsukov}},
  \bibinfo{author}{\bibfnamefont{Z.}~\bibnamefont{Frait}}, \bibnamefont{and}
  \bibinfo{author}{\bibfnamefont{W.}~\bibnamefont{Bailey}},
  \bibinfo{journal}{Phys.~Rev.~Lett.} \textbf{\bibinfo{volume}{98}},
  \bibinfo{pages}{117601} (\bibinfo{year}{2007}).

\bibitem[{\citenamefont{Steiauf and Faehnle}(2005)}]{Fahnle:2005}
\bibinfo{author}{\bibfnamefont{D.}~\bibnamefont{Steiauf}} \bibnamefont{and}
  \bibinfo{author}{\bibfnamefont{M.}~\bibnamefont{Faehnle}},
  \bibinfo{journal}{Phys.~Rev.~B} \textbf{\bibinfo{volume}{72}},
  \bibinfo{pages}{064450} (\bibinfo{year}{2005}).

\bibitem[{\citenamefont{Bhagat and Lubitz}(1974)}]{Bhagat.Lubitz:1974}
\bibinfo{author}{\bibfnamefont{S.}~\bibnamefont{Bhagat}} \bibnamefont{and}
  \bibinfo{author}{\bibfnamefont{P.}~\bibnamefont{Lubitz}},
  \bibinfo{journal}{Phys.~Rev.~B} \textbf{\bibinfo{volume}{10}},
  \bibinfo{pages}{179} (\bibinfo{year}{1974}).

\bibitem[{\citenamefont{Heinrich et~al.}(1979)\citenamefont{Heinrich, Meredith,
  and Cochran}}]{Heinrich:1979}
\bibinfo{author}{\bibfnamefont{B.}~\bibnamefont{Heinrich}},
  \bibinfo{author}{\bibfnamefont{D.}~\bibnamefont{Meredith}}, \bibnamefont{and}
  \bibinfo{author}{\bibfnamefont{J.}~\bibnamefont{Cochran}},
  \bibinfo{journal}{J.~Appl.~Phys.} \textbf{\bibinfo{volume}{50}},
  \bibinfo{pages}{7726} (\bibinfo{year}{1979}).

\bibitem[{\citenamefont{J.F.Cochran and Heinrich}(1980)}]{Heinrich:1980}
\bibinfo{author}{\bibnamefont{J.F.Cochran}} \bibnamefont{and}
  \bibinfo{author}{\bibfnamefont{B.}~\bibnamefont{Heinrich}},
  \bibinfo{journal}{IEEE Trans.~Magn.} \textbf{\bibinfo{volume}{16}},
  \bibinfo{pages}{660} (\bibinfo{year}{1980}).

\bibitem[{\citenamefont{Kambersky}(1976)}]{Kambersky:1976}
\bibinfo{author}{\bibfnamefont{V.}~\bibnamefont{Kambersky}},
  \bibinfo{journal}{Czech.~J.~Phys.~B} \textbf{\bibinfo{volume}{26}},
  \bibinfo{pages}{1366} (\bibinfo{year}{1976}).

\bibitem[{\citenamefont{Heinrich et~al.}(1967)\citenamefont{Heinrich, Fraitova,
  and Kambersky}}]{Kambersky:1967}
\bibinfo{author}{\bibfnamefont{B.}~\bibnamefont{Heinrich}},
  \bibinfo{author}{\bibfnamefont{D.}~\bibnamefont{Fraitova}}, \bibnamefont{and}
  \bibinfo{author}{\bibfnamefont{V.}~\bibnamefont{Kambersky}},
  \bibinfo{journal}{Phys.~Stat.~Sol.} \textbf{\bibinfo{volume}{23}},
  \bibinfo{pages}{501} (\bibinfo{year}{1967}).

\bibitem[{\citenamefont{Kambersky}(1970)}]{Kambersky:1970}
\bibinfo{author}{\bibfnamefont{V.}~\bibnamefont{Kambersky}},
  \bibinfo{journal}{Can.~J.~Phys.} \textbf{\bibinfo{volume}{48}},
  \bibinfo{pages}{2906} (\bibinfo{year}{1970}).

\bibitem[{\citenamefont{Kambersky}(1984)}]{Kambersky:1984}
\bibinfo{author}{\bibfnamefont{V.}~\bibnamefont{Kambersky}},
  \bibinfo{journal}{Czech.~J.~Phys.~B} \textbf{\bibinfo{volume}{34}},
  \bibinfo{pages}{1111} (\bibinfo{year}{1984}).

\bibitem[{\citenamefont{Korenman and Prange}(1972)}]{Korenman.Prange:1972}
\bibinfo{author}{\bibfnamefont{V.}~\bibnamefont{Korenman}} \bibnamefont{and}
  \bibinfo{author}{\bibfnamefont{R.}~\bibnamefont{Prange}},
  \bibinfo{journal}{Phys.~Rev.~B} \textbf{\bibinfo{volume}{6}},
  \bibinfo{pages}{2769} (\bibinfo{year}{1972}).

\bibitem[{\citenamefont{Kunes and Kambersky}(2002)}]{Kambersky:2002}
\bibinfo{author}{\bibfnamefont{J.}~\bibnamefont{Kunes}} \bibnamefont{and}
  \bibinfo{author}{\bibfnamefont{V.}~\bibnamefont{Kambersky}},
  \bibinfo{journal}{Phys.~Rev.~B} \textbf{\bibinfo{volume}{65}},
  \bibinfo{pages}{212411} (\bibinfo{year}{2002}).

\bibitem[{\citenamefont{Ingvarsson et~al.}(2002)\citenamefont{Ingvarsson,
  Ritchie, Liu, Xiao, Slonczewski, Trouilloud, and Koch}}]{Ingvarsson:2002}
\bibinfo{author}{\bibfnamefont{S.}~\bibnamefont{Ingvarsson}},
  \bibinfo{author}{\bibfnamefont{L.}~\bibnamefont{Ritchie}},
  \bibinfo{author}{\bibfnamefont{X.}~\bibnamefont{Liu}},
  \bibinfo{author}{\bibfnamefont{G.}~\bibnamefont{Xiao}},
  \bibinfo{author}{\bibfnamefont{J.}~\bibnamefont{Slonczewski}},
  \bibinfo{author}{\bibfnamefont{P.}~\bibnamefont{Trouilloud}},
  \bibnamefont{and} \bibinfo{author}{\bibfnamefont{R.}~\bibnamefont{Koch}},
  \bibinfo{journal}{Phys.~Rev.~B} \textbf{\bibinfo{volume}{66}},
  \bibinfo{pages}{214416} (\bibinfo{year}{2002}).

\bibitem[{\citenamefont{Mattheiss and Hamann}(1986)}]{Mattheiss.Hamann:1986}
\bibinfo{author}{\bibfnamefont{L.}~\bibnamefont{Mattheiss}} \bibnamefont{and}
  \bibinfo{author}{\bibfnamefont{D.}~\bibnamefont{Hamann}},
  \bibinfo{journal}{Phys.~Rev.~B} \textbf{\bibinfo{volume}{33}},
  \bibinfo{pages}{823} (\bibinfo{year}{1986}).

\bibitem[{\citenamefont{Hohenberg and Kohn}(1964)}]{Hohenberg.Kohn:1964}
\bibinfo{author}{\bibfnamefont{P.}~\bibnamefont{Hohenberg}} \bibnamefont{and}
  \bibinfo{author}{\bibfnamefont{W.}~\bibnamefont{Kohn}},
  \bibinfo{journal}{Phys.~Rev.} \textbf{\bibinfo{volume}{136}},
  \bibinfo{pages}{B864} (\bibinfo{year}{1964}).

\bibitem[{\citenamefont{Kohn and Sham}(1965)}]{Kohn.Sham:1965}
\bibinfo{author}{\bibfnamefont{W.}~\bibnamefont{Kohn}} \bibnamefont{and}
  \bibinfo{author}{\bibfnamefont{L.}~\bibnamefont{Sham}},
  \bibinfo{journal}{Phys.~Rev.} \textbf{\bibinfo{volume}{140}},
  \bibinfo{pages}{A1133} (\bibinfo{year}{1965}).

\bibitem[{\citenamefont{von Barth and Hedin}(1972)}]{Barth.Hedin:1972}
\bibinfo{author}{\bibfnamefont{U.}~\bibnamefont{von Barth}} \bibnamefont{and}
  \bibinfo{author}{\bibfnamefont{L.}~\bibnamefont{Hedin}},
  \bibinfo{journal}{J.~Phys.~C} \textbf{\bibinfo{volume}{5}},
  \bibinfo{pages}{1629} (\bibinfo{year}{1972}).

\bibitem[{\citenamefont{Stiles et~al.}(2001)\citenamefont{Stiles, Halilov,
  Hyman, and Zangwill}}]{Stiles:2001}
\bibinfo{author}{\bibfnamefont{M.}~\bibnamefont{Stiles}},
  \bibinfo{author}{\bibfnamefont{S.}~\bibnamefont{Halilov}},
  \bibinfo{author}{\bibfnamefont{R.}~\bibnamefont{Hyman}}, \bibnamefont{and}
  \bibinfo{author}{\bibfnamefont{A.}~\bibnamefont{Zangwill}},
  \bibinfo{journal}{Phys.~Rev.~B} \textbf{\bibinfo{volume}{64}},
  \bibinfo{pages}{104430} (\bibinfo{year}{2001}).

\bibitem[{\citenamefont{McMichael and Kunz}(2002)}]{McMichael:2002}
\bibinfo{author}{\bibfnamefont{R.}~\bibnamefont{McMichael}} \bibnamefont{and}
  \bibinfo{author}{\bibfnamefont{A.}~\bibnamefont{Kunz}},
  \bibinfo{journal}{J.~Appl.~Phys.} \textbf{\bibinfo{volume}{91}},
  \bibinfo{pages}{8650} (\bibinfo{year}{2002}).

\bibitem[{\citenamefont{Rossi et~al.}(2005)\citenamefont{Rossi, Heinonen, and
  MacDonald}}]{Heinonen:2005}
\bibinfo{author}{\bibfnamefont{E.}~\bibnamefont{Rossi}},
  \bibinfo{author}{\bibfnamefont{O.~G.} \bibnamefont{Heinonen}},
  \bibnamefont{and} \bibinfo{author}{\bibfnamefont{A.~H.}
  \bibnamefont{MacDonald}}, \bibinfo{journal}{Phys.~Rev.~B}
  \textbf{\bibinfo{volume}{72}}, \bibinfo{pages}{174412}
  (\bibinfo{year}{2005}).

\bibitem[{\citenamefont{Tserkovnyak et~al.}(2004)\citenamefont{Tserkovnyak,
  Fiete, and Halperin}}]{Tserkovnyak:2004}
\bibinfo{author}{\bibfnamefont{Y.}~\bibnamefont{Tserkovnyak}},
  \bibinfo{author}{\bibfnamefont{G.}~\bibnamefont{Fiete}}, \bibnamefont{and}
  \bibinfo{author}{\bibfnamefont{B.}~\bibnamefont{Halperin}},
  \bibinfo{journal}{Appl.~Phys.~Lett.} \textbf{\bibinfo{volume}{84}},
  \bibinfo{pages}{5234} (\bibinfo{year}{2004}).

\end{thebibliography}

\end{document}